\documentclass[
onecolumn
 reprint,
 amsmath,amssymb,
 aps,
]{revtex4-2}

\usepackage[linesnumbered, ruled]{algorithm2e}

\usepackage{xcolor}

\usepackage{enumitem}

\usepackage{graphicx}% Include figure files
\usepackage{dcolumn}% Align table columns on decimal point
\usepackage{comment}
\usepackage{bm}% bold math
\newcommand{\Nmax}{N_{\max}}

\begin{document}

\preprint{APS/123-QED}

\title{Accelerating Eigenvalue Computation for 
\\ Nuclear Structure Calculations via Perturbative Corrections
% Accelerating Nuclear Structure Calculations with Subspace Projection and Perturbative Corrections
% A new method for large-scale eigenvalue problems in nuclear structure calculations
}% Force line breaks with \\
% \thanks{A footnote to the article title}%

\author{Dong Min Roh}
\affiliation{Applied Mathematics and Computational Research Division, Lawrence Berkeley National Laboratory}%Lines break automatically or can be forced with \\
\author{Dean Lee}
\affiliation{Facility for Rare Isotope Beams and Department of Physics and Astronomy, Michigan State University, East Lansing, Michigan 48824}%
\author{Pieter Maris}
\affiliation{Department of Physics and Astronomy, Iowa State University, Ames, Iowa 50011}
\author{Esmond Ng}
\affiliation{Applied Mathematics and Computational Research Division, Lawrence Berkeley National Laboratory}
\author{James P. Vary}
\affiliation{Department of Physics and Astronomy, Iowa State University, Ames, Iowa 50011}
\author{Chao Yang}
\affiliation{Applied Mathematics and Computational Research Division, Lawrence Berkeley National Laboratory}

 % \email{Second.Author@institution.edu}
% \affiliation{%
%  Authors' institution and/or address\\
%  This line break forced with \textbackslash\textbackslash
% }%

% \collaboration{MUSO Collaboration}%\noaffiliation

% \author{Charlie Author}
%  \homepage{http://www.Second.institution.edu/~Charlie.Author}
% \affiliation{
%  Second institution and/or address\\
%  This line break forced% with \\
% }%
% \affiliation{
%  Third institution, the second for Charlie Author
% }%
% \author{Delta Author}
% \affiliation{%
%  Authors' institution and/or address\\
%  This line break forced with \textbackslash\textbackslash
% }%

% \collaboration{CLEO Collaboration}%\noaffiliation

% \date{\today}% It is always \today, today,
             %  but any date may be explicitly specified

\begin{abstract}

%Nuclear structure calculations are critical for understanding the properties of atomic nuclei. 
%The Configuration Interaction (CI) method simplifies the $A$-body Schrödinger equation to an eigenvalue problem, but solving this problem for large matrices is computationally expensive. 
We present a new method for computing the lowest few eigenvalues and the corresponding eigenvectors of a nuclear many-body Hamiltonian represented in a truncated configuration interaction subspace, i.e., the no-core shell model (NCSM). The method uses the hierarchical structure of the NCSM Hamiltonian to partition the Hamiltonian as the sum of two matrices.  The first matrix corresponds to the Hamiltonian represented in a small configuration space, whereas the second is viewed as the perturbation to the first matrix.  Eigenvalues and eigenvectors of the first matrix can be computed efficiently.  Perturbative corrections to the eigenvectors of the first matrix can be obtained from the solutions of a sequence of linear systems of equations defined in the small configuration space. These correction vectors can be combined with the approximate eigenvectors of the first matrix to construct a subspace from which more accurate approximations of the desired eigenpairs can be obtained. We call this method a 
Subspace Projection with Perturbative Corrections (SPPC) method. We show by numerical examples that the SPPC method can be more efficient than conventional iterative methods for solving large-scale eigenvalue problems such as the Lanczos, block Lanczos and the locally optimal block preconditioned conjugate gradient (LOBPCG) method. 
The method can also be combined with other methods to avoid
convergence stagnation. 
%Our numerical experiments on Hamiltonians of various nuclei demonstrate that the SPPC+RMM-DIIS significantly reduces the number of sparse matrix-vector multiplications (SpMVs) needed for convergence compared to traditional methods. 
%This hybrid approach offers a promising solution for large-scale nuclear structure calculations.

% \begin{description}
% \item[Usage]
% Secondary publications and information retrieval purposes.
% \item[Structure]
% You may use the \texttt{description} environment to structure your abstract;
% use the optional argument of the \verb+\item+ command to give the category of each item. 
% \end{description}
\end{abstract}

%\keywords{Suggested keywords}%Use showkeys class option if keyword
                              %display desired
\maketitle

%\tableofcontents

%%%%%%%%%%%%%%%%%%%%%%%%%%%%%%%%%%%%%%%%%%%%%%%%%%
\section{\label{sec:intro}Introduction}

% \begin{itemize}
%     \item Intro to MFDn and atomic nuclei Hamiltonian structure and its constructions..
%     \item Read \cite{cook2016high,maris2022accelerating}
% \end{itemize}

Nuclear structure calculations require solving $A$-body Schr\"{o}dinger equations where $A=Z+N$ is the number of nucleons consisting of $Z$ protons and $N$ neutrons. 
The Configuration Interaction (CI) method or no-core shell method (NCSM)~\cite{barrett2013ab}, which represents the solution to the Schr\"{o}dinger's equation by a linear combination of $A$-body basis functions, reduces the problem to an algebraic eigenvalue problem:
\begin{eqnarray}\label{eq:EP}
    H\Psi_k = E_k \Psi_k,
\end{eqnarray}
where $H\in\mathbb{R}^{n\times n}$ is the matrix representation of the $A$-body nuclear Hamiltonian operator in a configuration space (spanned by a set of $n$ $A$-body basis functions), $E_k$ is the $k$th eigenvalue of $H$ representing an approximation to a discrete energy level, and $\Psi_k$ is the corresponding eigenvector that contains the coefficients of the $A$-body basis function in the expansion of the approximate eigenfunction in the $A$-body basis.

The dimension ($n$) of the matrix $H$ depends on the number of nucleons $A$ and the size of the CI model space (determined by a truncation parameter $\Nmax$).
Although $n$ can be quite large, $H$ is very sparse, and often only a few of its eigenpairs at the low end of the spectrum are of interest, making iterative methods suitable for solving \eqref{eq:EP}.

\begin{figure}
\includegraphics[scale=0.4]{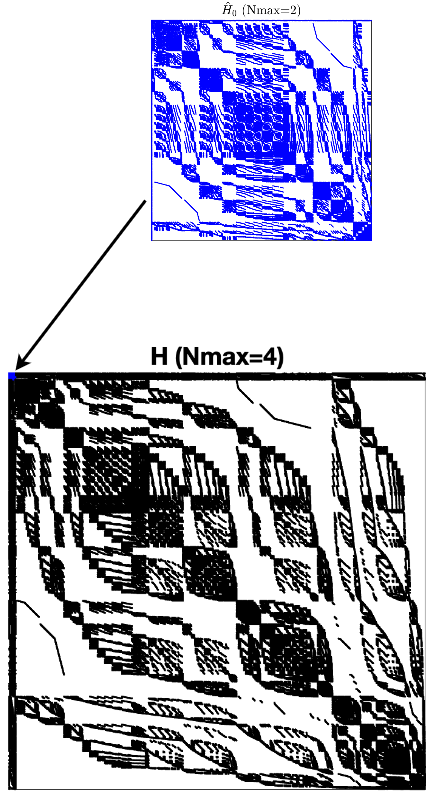}
\quad
\includegraphics[scale=0.25]{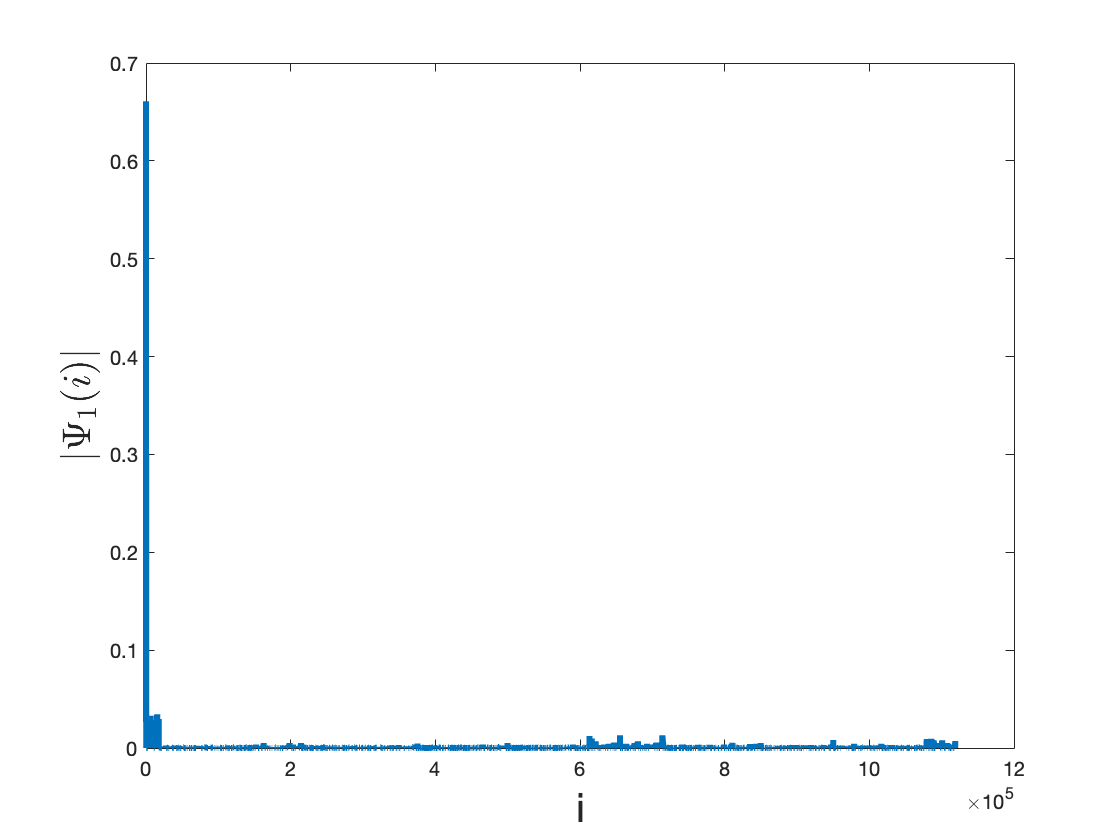}
\caption{\label{fig:C12}(a) The Hamiltonian matrix $H$ of ${}^{12}$C constructed by the CI method with a truncation parameter $\Nmax=4$, using the nucleon-nucleon interaction. The leading submatrix $\hat{H}_0$ of $H$ that is around $63$ times smaller is equivalent to the Hamiltonian matrix constructed by the CI method with a truncation parameter $\Nmax=2$. (b) Illustration of the eigenvector localization observed in the Hamiltonian matrix. The leading components of the eigenvector corresponding to the lowest eigenvalue of $H$ are several magnitudes larger than the tailing components.}
\end{figure}

The construction of the matrix $H$ by the CI method is typically done in a hierarchical fashion where the leading submatrix of $H$ corresponds to a matrix constructed from a smaller configuration space.
Because the $A$-body basis functions that form the lower dimension CI space associated with a small $\Nmax$ are typically more important than basis functions outside of such a configuration space, the eigenvectors of $H$ tend to be localized; i.e., the leading components of the $\Psi_k$ tend to be the larger in magnitude, while the tailing components are relatively small in magnitude.
We illustrate these properties of a nuclear Hamiltonian and its wavefunction using nucleus ${}^{12}$C as an example.
The Hamiltonian is constructed with a nucleon-nucleon interaction Daejeon16 \cite{shirokov2016n3lo} where hbar-omega value of $20$ describes the harmonic oscillator basis functions.
Figure~\ref{fig:C12}(a) shows that the leading submatrix of $H$ that is $63$ times smaller corresponds to a matrix constructed from a smaller configuration space.
Figure~\ref{fig:C12}(b) shows that the eigenvector corresponding to the lowest eigenvalue is localized in its first few components.
Therefore, the vector formed by padding the eigenvector of the leading submatrix with zeros can serve as a good initial guess for many iterative methods used to solve large-scale eigenvalue problems~\eqref{eq:EP}.
Previous works \cite{shao2018accelerating, alperen2023hybrid} select such initial guesses for algorithms like the Lanczos algorithm \cite{lanczos1950iteration}, the block Lanczos algorithm \cite{golub1977block}, the Locally Optimal Block Preconditioned Conjugate Gradient (LOBPCG) algorithm \cite{knyazev2001toward}, and the Residual Minimization Method with Direct Inversion of Iterative Subspace (RMM-DIIS) correction \cite{wood1985new,jia2001analysis,pulay1980convergence}.
Another study \cite{hernandez2021greedy} uses greedy algorithms to incrementally enlarge the submatrix and use the eigenvector of the enlarged submatrix as an improved starting guess.

% \begin{figure}[b]
% \includegraphics[scale=0.4]{C12_spy.png}
% \caption{\label{fig:C12_spy}The leading submatrix $\hat{H}_0$ (equivalent to a matrix generated by the CI method with a truncation parameter $\Nmax=2$) of the matrix $H_0$ (generated by the CI method with $\Nmax=4$) corresponding to the nucleus ${}^{12}$C is around $100$ times smaller.}
% \end{figure}

% \begin{figure}[b]
% \includegraphics[scale=0.2]{C12_smallest_eigenvector.png}
% \caption{\label{fig:C12_loc}Illustration of the eigenvector localization observed in the nucleus ${}^{12}$C. The leading components of the eigenvector corresponding to the smallest eigenvalue of $H$ are several magnitudes larger than other components.}
% \end{figure}

In this paper, we propose solving the eigenvalue problem~\eqref{eq:EP} with an iterative subspace projection method constructed from basis vectors obtained from eigenvalue and eigenvector perturbation theory. We call this method a Subspace Projection with Perturbative Corrections (SPPC) method. In this approach, the Hamiltonian matrix to be partially diagonalized is viewed as the sum of two matrices, i.e., $H=H_0+V$, where the eigenpairs of $H_0$ are relatively easy and inexpensive to compute, and $V$ is viewed as a perturbation to $H_0$. Perturbative corrections to eigenpairs of $H_0$ in successively higher order can be computed by solving a sequence of linear systems of equations. Together with the initial approximation to the desired eigenvector obtained from $H_0$, these correction vectors form a low dimensional subspace from which approximate eigenpairs of $H$ are extracted through the Rayleigh-Ritz procedure. When $H_0$ is chosen to be a block diagonal matrix diag($\hat{H}_0$,0), where $\hat{H}_0$ is the matrix representation of the $A$-body nuclear Hamiltonian in a smaller configuration space associated with a smaller $\Nmax$ value, these linear systems can be solved efficiently in the small configuration space.  The overall computational cost of the SPPC method grows linearly with respect to the highest order of perturbation included in the correction subspace.  Adding each perturbative correction vector and performing the Rayleigh-Ritz calculation requires multiplying the sparse matrix $H$ with a single vector. We show numerically that the low dimensional subspace constructed in a low order SPPC method provides more accurate approximation to a few lowest eigenvalues of $H$ than a Krylov subspace of the same dimension constructed from the same starting guess. The method also appears to be more efficient than the locally optimal block preconditioned conjugate gradient (LOBPCG) method in early iterations even when a good preconditioner is available for the LOBPCG method. Although the SPPC method can stagnate as higher perturbative corrections are included,  convergence stagnation can be mitigated by combining SPPC with other iterative algorithms for solving large scale eigenvalue problems using the SPPC's eigenvector approximation as the starting guess for secondary algorithms.

The rest of this paper is organized as follows. In Section~\ref{sec:sppc} we describe the basic Subspace Projection with Perturbative Corrections (SPPC) method for computing one eigenpair of $H$, and show how successively higher order perturbative corrections can be obtained from solutions of a set of linear systems of equations.
We draw the connection between SPPC and previously developed eigenvector continuation methods.
In Section~\ref{sec:several}, we present a version of the SPPC algorithm that can be used to compute a few eigenpairs.
A few practical implementation details of the SPPC method are discussed in Section~\ref{sec:details}.
Numerical examples that demonstrate the efficiency of the SPPC method relative to other conventional large-scale eigenvalue computation methods are presented in Section~\ref{sec:other}. We also show the effectiveness of combining SPPC with a conventional eigensolver for computing a few lowest eigenpairs of several nuclei.

% Finally, Section ~ concludes the paper and outlines potential future research directions.

% \CY{mention eigenvector localization and previous work on greedy algorithms.}

%DM - paragraph on EC
% Eigenvector continuation (EC) \cite{frame2018eigenvector} is a technique aimed at identifying the extremal eigenstates of a large-scale Hamiltonian $H(c)$, which smoothly varies with a parameter $c \in \mathbb{R}$. 
% The key idea behind EC is that despite their large dimension the trajectory of the exteremal eigenstates traces out a low-dimensional manifold as the parameter varies. 
% To approximate this manifold, EC selects its basis as the extremal eigenstates of the Hamiltonians at several parameter values where the computation is tractable. 
% This approach can be viewed as an analytic continuation that extends the expansion beyond its radius of convergence, allowing for the computation of eigenstates at parameter values where perturbative expansions fail \cite{frame2018eigenvector,duguet2023eigenvector}.

%%%%%%%%%%%%%%%%%%%%%%%%%%%%%%%%%%%%%%%%%%%%%%%%%%
\section{\label{sec:sppc}Subspace Projection with Perturbative Corrections (SPPC)}

We split the matrix $H$ into the sum of two matrices $H_0$ and $V$, i.e.,
\begin{eqnarray}\label{eq:split}
    H = H_0 + V,
\end{eqnarray}
where 
\begin{eqnarray}\label{eq:split2}
    H_0 := \begin{bmatrix}
        \hat{H}_0 & 0 \\
        0 & 0
    \end{bmatrix}
    , \quad
    V := \begin{bmatrix}
        0 & V_{12} \\
        V_{21} & V_{22}
    \end{bmatrix},
\end{eqnarray}
with the matrix $\hat{H}_0\in\mathbb{R}^{n_0\times n_0}$ (where $n_0\ll n$) being the leading submatrix of $H$ that corresponds to the representation of the Hamiltonian within a smaller configuration space.

The eigenvectors of $H_0$, which can be obtained from the eigenvectors of $\hat{H}_0$, are computed with a much lower computational cost compared to those of the full matrix $H$. They can be used as good initial guesses for conventional algorithms, such as the Lanczos and the LOBPCG algorithms, for computing the eigenpairs of $H$.

The SPPC method uses the eigenvectors of $H_0$ to initiate a subspace construction procedure to produce a subspace from which improved approximations of the desired eigenpairs can be obtained.

Instead of using the Lanczos algorithm or the LOBPCG method to construct the subspace, we use perturbative corrections to the initial eigenvector approximation to construct the subspace. In this approach, we view $H$ as a matrix obtained from perturbing $H_0$ by $cV$ with $c=1$, 
i.e., we can write $H$ as
\begin{eqnarray}\label{eq:Hc}
    H(c) = H_0 + cV.
\end{eqnarray}
%with $c=1$.

It follows from the Rayleigh-Schr\"{o}dinger perturbation theory \cite{shavitt2009many} that the eigenvalues and eigenvectors of $H(c)$, which we denote by $E_k(c)$ and $\Psi_k(c)$ respectively, can be written in terms of perturbative corrections to the eigenvalues and eigenvectors of $H_0$, which we denote by
$E_k^{(0)}$ and $\Psi_k^{(0)}$, i.e.,
\begin{eqnarray}\label{eq:exp}
    \begin{cases}
        E_k(c) &= E_k^{(0)} + c E_k^{(1)} + c^2 E_k^{(2)} + \cdots, \\
        \Psi_k(c) &= \Psi_k^{(0)} + c \Psi_k^{(1)} + c^2 \Psi_k^{(2)} + \cdots.
    \end{cases}
\end{eqnarray}
Here, ${(E_k^{(p)}, \Psi_k^{(p)})}_{p\geq1}$ represent the $p$th order perturbative corrections to the eigenpair $(E_k^{(0)}, \Psi_k^{(0)})$, and are independent of the parameter $c$.

Substituting \eqref{eq:exp} into the equation
\begin{eqnarray}\label{eq:EPc}
    H(c)\Psi_k(c) = E_k(c) \Psi_k(c)
\end{eqnarray}
and matching coefficients of the same degree order
yields the following set of equations 
\begin{eqnarray}\label{eq:exp_seq}
    \begin{cases}
        E_k^{(p)} = (\Psi_k^{(p-1)})^T V \Psi_k^{(0)} \\
        (H_0 - E_k^{(0)}) \Psi_k^{(p)} = (E_k^{(1)} - V)\Psi_k^{(p-1)} + \sum_{l=0}^{p-2}E_k^{(p-l)}\Psi_k^{(l)}
    \end{cases}
\end{eqnarray}
that allow us to compute $E_k^{(p)}$ and $\Psi_k^{(p)}$ in a recursive fashion.

The asymptotic expansion used in \eqref{eq:exp} assumes that $c$ is a small parameter. As a result, the expansion serves as a good approximation to the desired eigenpair only when $c$ is sufficiently small, i.e., when $c$ falls within the radius of convergence for \eqref{eq:exp}, which is generally much smaller than 1.  As a result, \eqref{eq:exp} cannot be used directly in general to approximate the $k$th eigenpair of $H$ \cite{franzke2022excited,duguet2023eigenvector}. However, the perturbative vectors $\Psi_k^{(p)}$ can be used to construct a subspace
\begin{eqnarray}\label{eq:Man_P}
    \mathcal{M}_k^{(P)} := span\{\Psi_k^{(p)}:p=0,1,\ldots,P\},
\end{eqnarray}
from which approximation to $E_k$ and $\Psi_k$ can be obtained. 

The idea of using perturbative corrections to construct an approximating subspace was proposed in~\cite{franzke2022excited,demol2020improved} in the context of an eigenvector continuation (EC) method~\cite{frame2018eigenvector,sarkar2021convergence}. In an EC method, the eigenvectors of $H(c)$ for some choices of $c$'s are used to construct a subspace from which approximations to the eigenvectors of $H(c')$ for $c'\neq c$ are obtained from the projection of $H(c')$ into such a subspace.   

It was found in ~\cite{franzke2022excited,demol2020improved} that instead of using eigenvectors of $H(c)$ for several choices of $c\neq 1$ to construct a subspace from which approximate eigenvectors of $H(1)$ are extracted through the standard Rayleigh-Ritz procedure, more accurate approximations to the desired eigenpairs of $H(1)$ can be obtained from the subspace constructed from the eigenvector of $H_0$ as well as perturbative eigenvector corrections as discussed above.

In a Rayleigh-Ritz procedure, we compute an orthonormal basis matrix $Q_k^{(P)}$ of $\mathcal{M}_k^{(P)}$ and form the projected matrix
\begin{equation}\label{eq:tilde_H}
\tilde{H} = \left(Q_k^{(P)}\right)^T H Q_k^{(P)}.
\end{equation}
If $(\theta, v)$ is an eigenpair of $\tilde{H}$, then $(\theta,z)$ where $z = Q_k^{(P)}v$, yields an approximate eigenpair of $H$.
We consider an approximate eigenpair to have converged if its relative residual norm
\begin{eqnarray}\label{eq:rrn}
    \frac{\|Hz - \theta z\|_2}{|\theta|}
\end{eqnarray}
is smaller than a preset tolerance value. 

To obtain an approximate $k$th eigenpair $(\tilde{E}_k, \tilde{\Psi}_k)$ of $H$, we choose $(\theta,v)$ as the \textit{lowest} eigenpair of $\tilde{H}$ in the above mentioned Rayleigh-Ritz procedure.
We present the SPPC method in Algorithm~\ref{alg:SPPC_k}.

\begin{algorithm}\label{alg:SPPC_k}
    \DontPrintSemicolon 

    \KwIn{A nuclear CI Hamiltonian $H\in\mathbb{R}^{n\times n}$ partitioned as $H=H_0 + V$, where $H_0=diag(\hat{H}_0,0)$ with $\hat{H}_0$ constructed from a small configuration space (of dimension $n_0$); convergence tolerance ($tol$); and maximum order of perturbation allowed ($maxiter$)}
    \KwOut{An approximate $k$th eigenpair $(\tilde{E}_k, \tilde{\Psi}_k)$ of $H$}

    %Split $H$ according to \eqref{eq:split} and \eqref{eq:split2}.
    
    Compute the $k$th nonzero eigenpair $(E_k^{(0)},\Psi_k^{(0)})$ of $H_0$.

    Set $\theta=(\Psi_k^{(0)})^TH\Psi_k^{(0)}$ and $z=\Psi_k^{(0)}$.

    Return $(\tilde{E}_k=\theta, \tilde{\Psi}_k=z)$ if the relative residual norm~\eqref{eq:rrn} is less than $tol$.

    \For{$p=1,\ldots,maxiter$}{

        Compute the correction energy $E_k^{(p)}$ and correction vector $\Psi_k^{(p)}$.

        Compute an orthonormal basis matrix $Q_k^{(p)}$ of $\mathcal{M}_k^{(p)}$ and form a projected matrix $\tilde{H}$.

        Compute the lowest eigenpair $(\theta,q)$ of $\tilde{H}$ and set $z=Q_k^{(p)}q$.
    
        Return $(\tilde{E}_k=\theta, \tilde{\Psi}_k=z)$ if the relative residual norm~\eqref{eq:rrn} is less than $tol$.

    }

    \caption{The SPPC for the $k$th eigenpair}

\end{algorithm}

%
%It draws motivation from a technique called eigenvector continuation (EC) \cite{frame2018eigenvector}. 
%EC identifies the eigenvectors of a large-scale Hamiltonian $H(c)$ that smoothly varies with a parameter $c\in\mathbb{R}$.
%It first constructs a subspace whose basis consists of the eigenvectors of the Hamiltonians at several parameter values where the computation is tractable, then performs the Rayleigh-Ritz procedure with this subspace by projecting the Hamiltonian of interest to the subspace and solving the eigenvalue problem of the projected matrix.
%To align with the idea of EC, we transform the matrix $H$ in \eqref{eq:split} into a parameter-dependent Hamiltonain
%
%such that the Hamiltonian at $c=1$ corresponds to the original matrix $H$.

%The perturbation expansions~\eqref{eq:exp} diverge for parameter values beyond the radius of convergence, especially for $c=1$.
%However, we can utilize the perturbative correction vectors to construct a subspace and perform the Rayleigh-Ritz procedure.
%Specifically, we construct the subspace

%\bigskip
%\bigskip

We should point out that the main distinction between the SPPC method proposed here and the EC method lies in the subspaces constructed in these methods and the cost of construction. Because EC projects $H$ onto a subspace constructed from the eigenvectors of $H(c)$ for several (nonzero) $c$'s, the cost of subspace construction may be just as expensive as the cost of solving the target eigenvalue problem with a particular choice of $c$. On the other hand, because the subspace constructed in the SPPC method uses perturbative corrections that can be obtained by solving much smaller linear systems, the cost of subspace construction is significantly lower.

%Both the SPPC and EC construct a subspace and perform the Rayleigh-Ritz procedure.
%The difference lies in their subspaces and the computational advantage of the SPPC in subspace construction.
%EC computes eigenvectors of full-sized Hamiltonian matrices $H(c)$ at several parameter points. 
%Each of these eigenvector computations can be just as expensive as solving for the eigenvector of the original matrix $H$.
%Additionally, the performance of EC depends on the number and location of the parameter points.
%Determining these hyperparameters can be challenging, often requiring domain knowledge or assistance from machine learning techniques \cite{sarkar2022self}.
%On the other hand, the SPPC is an iterative method that computes a single eigenvector $\Psi_k^{(0)}$ initially and successively computes the correction vectors $\Psi_k^{(p)}$.
%We show in Section~\ref{sec:details} that these computations can be done efficiently without having to work with full-sized matrices by leveraging a splitting scheme.

Although the basic idea of SPPC was presented in ~\cite{franzke2022excited,demol2020improved}, the computational cost of this method was not carefully analyzed and compared with those of the state-of-the-art large-scale eigensolvers. 
In~\cite{franzke2022excited}, the eigenvectors of $H_0$ can be computed analytically for the one-dimensional quartic anharmonic oscillator. However, in general, identifying a $H_0$ that can be diagonalized analytically is not possible. In~\cite{demol2020improved}, SPPC is used to perform a $A$-body nuclear structure calculation. However, a different $H =H_0+V$ splitting scheme is used, and the eigenvectors of $H_0$ are not easier to compute than those of $H$.

%%%%%%%%%%%%%%%%%%%%%%%%%%%%%%%%%%%%%%%%%%%%%%%%%%
\section{\label{sec:several}Targeting First Few Eigenpairs}

Although we can use Algorithm~\ref{alg:SPPC_k} to compute each of the first $k_{ev}$ eigenpairs one by one (or in parallel), approximations to larger eigenvalues and the corresponding eigenvectors appear to converge slowly, and sometimes to wrong values, as we will show in section~\ref{sec:num}. 
A more effective way to obtain approximations to the first $k_{ev}$ eigenpairs is to combine the SPPC subspace $\mathcal{M}_k^{(P)}$~\eqref{eq:Man_P} constructed for each eigenpair to create a larger subspace 
\begin{eqnarray}\label{eq:comb_sub}
    \mathcal{M}^{(P)} := \bigcup_{k=1}^{k_{ev}}\mathcal{M}_k^{(P)},
\end{eqnarray}
from which $k_{ev}$ approximate eigenpairs can be extracted simultaneously through the Rayleigh-Ritz procedure, by targeting the $k_{ev}$ lowest eigenpairs of the projected matrix.
Algorithm~\ref{alg:SPPC} shows how this approach works.

%\begin{enumerate}[label=(\alph*)]
%
%    \item We use the SPPC separately for each of the first $k_{ev}$ eigenpairs.
%    That is, we use Algorithm~\ref{alg:SPPC_k} for $k=1,\ldots,k_{ev}$.
%
%    % For $k=1,2,\ldots,k_{ev}$, we compute an orthonormal matrix $Q_k^{(P)}$ for each $\mathcal{M}_k^{(P)}$ and perform the Rayleigh-Ritz procedure by solving the eigenvalue problem of the projected matrix $\tilde{H}_k^{(P)}:=(Q_k^{(P)})^THQ_k^{(P)}$, targeting its smallest eigenpair.
%
%    \item We combine the subspaces $\mathcal{M}_k^{(P)}$~\eqref{eq:Man_P} for the first $k_{ev}$ eigenvectors:
%    \begin{eqnarray}\label{eq:comb_sub}
%        \mathcal{M}^{(P)} := \bigcup_{k=1}^{k_{ev}}\mathcal{M}_k^{(P)},
%    \end{eqnarray}
%    forming a subspace spanned by the correction vectors of all $k_{ev}$ eigenvectors.
%    Then, we compute an orthonormal matrix $Q^{(P)}$ of $\mathcal{M}^{(P)}$ and form the projected matrix
%    \begin{eqnarray}\label{eq:tilde_H2}
%        \tilde{H} = (Q^{(P)})^THQ^{(P)}.
%    \end{eqnarray}
%    To obtain an approximate $k$th eigenpair $(\tilde{E}_k,\tilde{\Psi}_k)$, we compute the $k$th eigenpair $(\theta_k,q_k)$ of $\tilde{H}$~\eqref{eq:tilde_H2} in a Rayleigh-Ritz procedure.    
%\end{enumerate}

\begin{algorithm}\label{alg:SPPC}
    \DontPrintSemicolon 

    \KwIn{A nuclear CI Hamiltonian $H\in\mathbb{R}^{n\times n}$ partitioned as $H=H_0 + V$, where $H_0=diag(\hat{H}_0,0)$ with $\hat{H}_0$ constructed from a small configuration space (of dimension $n_0$); number of desired eigenpairs ($k_{ev}$); convergence tolerance ($tol$); and maximum order of perturbation allowed ($maxiter$)}

    \KwOut{Approximate $k_{ev}$ lowest eigenpairs $\{(\tilde{E}_k, \tilde{\Psi}_k)\}_{k=1}^{k_{ev}}$ of $H$}

    Compute the eigenpairs $\{(E_k^{(0)}, \Psi_k^{(0)}\}_{k=1}^{k_{ev}}$ of $H_0$. 

    Compute an orthonormal basis matrix $Q^{(0)}$ of $\mathcal{M}^{(0)}$ and form the projected matrix $\tilde{H}$.

    Compute the $k_{ev}$ lowest eigenpairs $\{(\theta_k,q_k)\}_{k=1}^{k_{ev}}$ of $\tilde{H}$ and set $\{z_k=Q^{(0)}q_k\}_{k=1}^{k_{ev}}$.

    Return $\{(\tilde{E}_k=\theta_k, \tilde{\Psi}_k=z_k)\}_{k=1}^{k_{ev}}$ if the relative residual norm~\eqref{eq:rrn} is less than $tol$ for all $k=1,\ldots,k_{ev}$.
    
    \For{$p=1,\ldots,maxiter$}{
    
        Compute the correction energy $E_k^{(p)}$ and correction vector $\Psi_k^{(p)}$ for $k=1,\ldots,k_{ev}$.

        Compute an orthonormal basis matrix $Q^{(p)}$ of $\mathcal{M}^{(p)}$ and form a projected matrix $\tilde{H}$.

        Compute the $k_{ev}$ lowest eigenpairs $\{(\theta_k,q_k)\}_{k=1}^{k_{ev}}$ of $\tilde{H}$ and set $\{z_k=Q^{(p)}q_k\}_{k=1}^{k_{ev}}$.
    
        Return $\{(\tilde{E}_k=\theta_k, \tilde{\Psi}_k=z_k)\}_{k=1}^{k_{ev}}$ if the relative residual norm~\eqref{eq:rrn} is less than $tol$ for all $k=1,\ldots,k_{ev}$.

    }

    \caption{The SPPC for the first few eigenpairs}
\end{algorithm}

% We should note that it may be possible to use Algorithm~\ref{alg:SPPC_k} to compute a particular interior eigenvalue and the corresponding eigenvector without computing other eigenpairs. However, the algorithm may need to be combined with another method when the SPPC approximation fails to converge to the desired accuracy as we will discuss in section~\ref{sec:other}. 

%%%%%%%%%%%%%%%%%%%%%%%%%%%%%%%%%%%%%%%%%%%%%%%%%%
\section{\label{sec:details}Practical considerations}
In this section, we describe a few practical implementation details of the SPPC algorithm.
\subsection{\label{subsec:split_vecsolve} Computing the Correction Vectors}

%The splitting scheme~\eqref{eq:split2} allows us to efficiently obtain the eigenvector 
We refer to the $k$th eigenvector of $H_0$ 
 denoted by $\Psi_k^{(0)}$ as the zero-th order correction to the $k$th eigenvector of $H$.

 When $H_0$ is of the form given in \eqref{eq:split2}, $\Psi_k^{(0)}$ can be obtained by computing the $k$th eigenvector of $\hat{H}_0$, denoted by $\hat{\Psi}_k^{(0)}$, and appending it with zeros to yield
\begin{equation}\label{eq:EP_small}
    %\hat{H}_0\hat{\Psi}_k^{(0)}=E_k^{(0)}\hat{\Psi}_k^{(0)}, \quad 
   \Psi_k^{(0)} = \begin{bmatrix}
   \hat{\Psi}_k^{(0)} \\
    0
    \end{bmatrix}.
\end{equation}

It follows from \eqref{eq:exp_seq} and the block structures of $H_0$, $V$, and ${\Psi}_k^{(0)}$ that the first order corrections to the $k$th eigenvalue and eigenvector of $H$ are
\begin{eqnarray}\label{eq:first_PC}
    E_k^{(1)} = 0, \quad \Psi_k^{(1)} = \frac{1}{E_k^{(0)}}V\Psi_k^{(0)}.
\end{eqnarray}

It is easy to verify that 
\[
\Psi_k^{(1)} = \frac{H\Psi_k^{(0)} - E_k^{(0)}\Psi_k^{(0)}}{E_k^{(0)}},
\]
i.e., the first order correction to the eigenvector is simply the residual associated with the zero-th order approximation.

Because $E_k^{(1)}=0$, we can simplify the linear system in~\eqref{eq:exp_seq} to
\begin{eqnarray}\label{eq:exp_seq2}
    (H_0 - E_k^{(0)}) \Psi_k^{(p)} = -V\Psi_k^{(p-1)} + \sum_{l=0}^{p-2}E_k^{(p-l)}\Psi_k^{(l)}.
\end{eqnarray}
The matrix $H_0 - E_k^{(0)}$ is in a block diagonal form
\begin{eqnarray}\label{eq:bl_diag}
    H_0 - E_k^{(0)} = 
    \begin{bmatrix}
        \hat{H}_0 - E_k^{(0)} & 0 \\
        0 & -E_k^{(0)}I
    \end{bmatrix}
\end{eqnarray}
consisting of two blocks where the first block $\hat{H}_0 - E_k^{(0)}\in\mathbb{R}^{n_0\times n_0}$ is relatively small and the second block $-E_k^{(0)}I\in\mathbb{R}^{(n-n_0)\times(n-n_0)}$ is a scalar multiple of the identity matrix.
As a result, the solution of the linear system in \eqref{eq:exp_seq2} essentially reduces to the solution of a much smaller linear system with $\hat{H}_0 - E_k^{(0)}$ being the coefficient matrix.
If we partition $\Psi_k^{(p)}$ conformally with the blocks in \eqref{eq:bl_diag} as $\Psi_k^{(p)}=[x_1, x_2]^T$ and the right-hand side of \eqref{eq:exp_seq2} as $b=[b_1, b_2]^T$ such that $x_1,b_1\in\mathbb{R}^{n_0}$ and $x_2,b_2\in\mathbb{R}^{n-n_0}$,  $x_2$ can be easily computed as 
\begin{eqnarray}\label{eq:sm_first}
    x_2=-\frac{1}{E_k^{(0)}}b_2,
\end{eqnarray}
and $x_1$ can be obtained by solving
\begin{eqnarray}\label{eq:sm_lin}
    (\hat{H}_0-E_k^{(0)})x_1 = b_1.
\end{eqnarray}

Note that equation \eqref{eq:exp_seq2} is singular because $E_k^{(0)}$ is an eigenvalue of $H_0$.  However, since $\mathcal{M}_k^{(P)}$ already includes $\Psi_k^{(0)}$ and we are only interested in contributions in the orthogonal complement of $\Psi_k^{(0)}$ from the solution of \eqref{eq:exp_seq2}, we can project out  $\Psi_k^{(0)}$ from the right-hand side of \eqref{eq:exp_seq2} before solving this equation.
This is equivalent to projecting out $\hat{\Psi}_k^{(0)}$ from the right-hand side of \eqref{eq:sm_lin}, i.e. we solve
\begin{eqnarray}\label{eq:sm_lin_proj}
    (\hat{H}_0-E_k^{(0)})x_1 = b_1 - (b_1^T \hat{\Psi}_k^{(0)}) \hat{\Psi}_k^{(0)}.
\end{eqnarray}

\subsection{\label{subsec:RR} Rayleigh-Ritz Calculation}

After obtaining the correction vectors, we generate an orthonormal basis matrix of the subspace spanned by these vectors and then perform the Rayleigh-Ritz procedure.
We lay out these steps for approximating a single eigenpair and the first few eigenpairs.

\paragraph{Targeting the $k$th eigenpair.}
We use the Gram-Schmidt process to obtain an orthonormal basis  of the subspace $\mathcal{M}_k^{(p)}$. 
The orthonormal basis forms the columns of the matrix $Q_k^{(p)}\in\mathbb{R}^{n\times (p+1)}$. The $(p+1)$th column, denoted by $q_k$, is generated as follows.
\begin{equation}\label{eq:gs}
    \begin{split}
        \Phi_k^{(p)} &= \left[I - Q_k^{(p-1)}(Q_k^{(p-1)})^T\right]\Psi_k^{(p)}, \\
        q_k^{(p)} &= \frac{\Phi_k^{(p)}}{\|\Phi_k^{(p)}\|_2}.
    \end{split}
\end{equation}
We append $q_k^{(p)}\in\mathbb{R}^n$ to $Q_k^{(p-1)}$ such that
\begin{eqnarray}\label{eq:Qkp}
    Q_k^{(p)} = [Q_k^{(p-1)}, q_k^{(p)}].
\end{eqnarray}

Note that the projected matrix $(Q_k^{(p)})^THQ_k^{(p)}$ can be constructed recursively.  Assuming $(Q_k^{(p-1)})^THQ_k^{(p-1)}$ has been computed in the previous step, we just need to compute $Hq_k^{(p)}$ and append an additional row and column to $(Q_k^{(p-1)})^THQ_k^{(p-1)}$ as shown below.
\begin{equation}\label{eq:proj}
    \begin{bmatrix}
        (Q_k^{(p-1)})^THQ_k^{(p-1)} & (Q_k^{(p-1)})^THq_k^{(p)} \\
        (q_k^{(p)})^THQ_k^{(p-1)} & (q_k^{(p)})^THq_k^{(p)}
    \end{bmatrix}.
\end{equation}
Therefore, the major cost for constructing the projected matrix in each step of the SPPC method is in performing a single SpMV in $Hq_k^{(p)}$.

\paragraph{Targeting the first few eigenpairs.}
To obtain an orthonormal basis $Q^{(p)}\in\mathbb{R}^{n\times k_{ev}(p+1)}$ for the combined subspace $\mathcal{M}^{(p)}$, we replace the Gram-Schmidt process~\eqref{eq:gs} with the block Gram-Schmidt process.
If $\Psi^{(p)}=[\Psi_1^{(p)}, \Psi_2^{(p)}, \ldots, \Psi_{k_{ev}}^{(p)}]$, the block Gram-Schmidt procedure yields 
\begin{equation}\label{eq:blkGS}
        \Phi^{(p)} = \left[I - Q^{(p-1)}(Q^{(p-1)})^T\right]\Psi^{(p)}.
\end{equation}
We then perform a QR factorization of $\Phi^{(p)}$, i.e.,
\begin{equation}\label{eq:QR}
        \Phi^{(p)} = q^{(p)}R^{(p)},
\end{equation}
to generate an orthonormal basis $q^{(p)}$ for $\Phi^{(p)}$.
We append $q^{(p)}\in\mathbb{R}^{n\times k_{ev}}$ to $Q^{(p-1)}$ such that
\begin{eqnarray}\label{eq:Qp}
    Q^{(p)} = [Q^{(p-1)}, q^{(p)}]
\end{eqnarray}
is an orthonormal basis for $\mathcal{M}^{(p)}$.

Again, the projected matrix $(Q^{(p)})^THQ^{(p)}$ can be constructed recursively, and
%\begin{equation}\label{eq:proj2}
% (Q^{(p)})^THQ^{(p)} =    \begin{bmatrix}
%        (Q^{(p-1)})^THQ^{(p-1)} & (Q^{(p-1)})^THq^{(p)} \\
%        (q^{(p)})^THQ^{(p-1)} & (q^{(p)})^THq^{(p)}
%%    \end{bmatrix}.
%\end{equation}
the computational cost is dominated by the cost for computing $k_{ev}$ SpMVs in $Hq^{(p)}$.

\subsection{\label{subsec:alg+cost} Computational Cost}

We now discuss the overall computational cost of the SPPC method.  We can see from Algorithm~\ref{alg:SPPC_k} and Algorithm~\ref{alg:SPPC} that the two major components of the SPPC algorithm are: (1) Solving the linear system \eqref{eq:exp_seq2}; (2) forming the projected matrix 
$(Q^{(p)})^THQ^{(p)}$. We have already shown that the projected matrix can be computed recursively using $k_{ev}$ SpMVs in each step of the SPPC algorithm. The reduced linear system~\eqref{eq:sm_lin} of the correction equation \eqref{eq:exp_seq2} can be solved iteratively using, for example, the MINRES algorithm. Because it has a much smaller dimension, the SpMVs performed in each MINRES iteration are relatively cheap.  However, forming the right-hand side of the equation~\eqref{eq:exp_seq2} requires multiplying $V$ with $\Psi_k^{(p-1)}$ for $p>1$ which has nearly the same complexity as multiplying $H$ with $\Psi_k^{(p-1)}$. Therefore, it may appear that each SPPC step requires performing $2k_{ev}$ SpMVs. We will show below that this is not the case. Both the right-hand side of \eqref{eq:exp_seq2} and the projected matrix can be obtained from the same $Hq^{(p-1)}$ product. As a result, each step of the SPPC algorithm only requires performing $k_{ev}$ SpMVs.

Using the matrix splitting $H=H_0 + V$, we can rewrite $V\Psi^{(p-1)}$ as 
\begin{equation}\label{eq:avoid_SpMV2}
V\Psi^{(p-1)} = H\Psi^{(p-1)} - H_0\Psi^{(p-1)}
    \end{equation}
where $\Psi^{(p-1)}=\left[\Psi_1^{(p-1)}, \ldots, \Psi_{k_{ev}}^{(p-1)}\right]$. Therefore, $V\Psi^{(p-1)}$ can be obtained by subtracting $H_0\Psi^{(p-1)}$, a much lower computational cost, from $H\Psi^{(p-1)}$.

We now show that $H\Psi^{(p-1)}$ can be easily obtained from $Hq^{(p-1)}$.  It follows from \eqref{eq:blkGS} and \eqref{eq:QR} that
\begin{equation}\label{eq:HPsip}
H\left[I -Q^{(p-2)}(Q^{(p-2)})^T\right]\Psi^{(p-1)} = Hq^{(p-1)}R^{(p-1)} .
%        &&= HQ^{(p-1)}
%        \begin{bmatrix}
%            (Q^{(p-2)})^T\Psi^{(p-1)} \\
%            R^{(p-1)}
%        \end{bmatrix},
    \end{equation}
As a result, we can obtain $H\Psi^{(p-1)}$ from $Hq^{(p-1)}$ by using following identity
\begin{equation}
H\Psi^{(p-1)} = HQ^{(p-2)}(Q^{(p-2)})^T\Psi^{(p-1)} + Hq^{(p-1)}R^{(p-1)}.
\end{equation}
Note that the analysis of the computational cost assumes $HQ^{(p-2)}$,  which contains $Hq^{(j)}$ as its columns for $j = 0, 1, ..., p-2$, has been stored in memory.

As we will show in section~\ref{sec:num}, the highest order perturbation is often limited to 15, beyond which no significant improvement in the approximate eigenpair can be observed. Therefore, the dense linear algebra operations such as computing $(Q^{(p-2)})^T \Psi^{(p-1)}$ and diagonalizing the projected matrix can be performed with a relatively low cost compared to the cost of multiplying $H$ with $q^{(p-1)}$.

\section{\label{sec:other}Combining SPPC with other algorithms}
As we will show in the next section, the perturbative correction is typically effective when the order of perturbation $p$ is relatively low. The convergence of SPPC can stagnate 
 when $p$ increases, i.e., adding higher order perturbative correction may not help because they may be linearly dependent with respect to the basis vectors included in $\mathcal{M}^{(p)}$ already.  In this case, it is useful to combine SPPC with another algorithm that can take the eigenvector approximation produced by SPPC as the starting guess. 

Algorithms that can be combined with the SPPC method in a hybrid algorithm include but are not limited to the following:
\begin{itemize}

    \item The Lanczos algorithm, which is a classical algorithm that generates an orthonormal basis of a Krylov subspace using the Gram-Schmidt procedure.
    It is initialized with an approximate eigenvector produced from SPPC or a linear combination of $k_{ev}$ approximate eigenvectors. It uses one SpMV per iteration to compute the next basis vector.
    The eigenpairs are approximated using Ritz pairs obtained from the Rayleigh-Ritz procedure with the basis vectors of the Krylov subspace.
    
    \item The block Lanczos algorithm, which is a variation of the Lanczos algorithm that operates in blocks.
    It is initialized with a block of vectors approximating several eigenvectors of $H$, and builds a Krylov subspace in blocks, with each iteration performing $k_{ev}$ SpMVs.
    The main advantage of the block Lanczos algorithm over the Lanczos algorithm is that it can make use of approximations to several eigenvectors more effectively and most dense linear algebra operations can take advantage of level 3 BLAS.
    
    \item The LOBPCG algorithm, which is an iterative method that solves the equivalent trace minimization formulation of the eigenvalue problem.
    Similar to the block Lanczos algorithm, it can be initialized with approximations to several eigenvectors and each iteration performs $k_{ev}$ SpMVs.
    One advantage of the LOBPCG algorithm is that it can utilize a preconditioner if it is available. The use of a preconditioner can accelerate convergence. 
    A common choice for the preconditioner is a block diagonal part of the Hamiltonian matrix or a shifted matrix of the Hamiltonian for some appropriately chosen shift \cite{knyazev2001toward,alperen2023hybrid}.
    For our numerical experiments, we use a shifted preconditioner that involves a specific block diagonal part of the Hamiltonian matrix and a constant shift that approximates the lowest eigenvalue.
    
    \item The residual minimization method (RMM) combined with the Direct Inversion of Iterative Subspace (DIIS) refinement algorithm (RMM-DIIS), which is a quasi-Newton algorithm for improving  a specific eigenpair without computing other eigenpairs, provided that the initial approximation to the desired eigenpairs is sufficiently accurate. Each RMM-DIIS iteration performs one SpMV.  The RMM-DIIS can also incorporate a preconditioner to accelerate convergence. For our numerical experiments, we choose a preconditioner that is a specific diagonal part of the Hamiltonian matrix.
\end{itemize}

%%%%%%%%%%%%%%%%%%%%%%%%%%%%%%%%%%%%%%%%%%%%%%%%%%
\section{\label{sec:num}Numerical Results}

In this section, we demonstrate the effectiveness of SPPC algorithm and compare it with other existing algorithms for computing the ground and a few low excited states of several light nuclei. We also show that SPPC can be effectively combined with RMM-DIIS to yield an efficient and accurate hybrid eigensolver 
for nuclear configuration interaction calculations. We call this hybrid eigensolver the SPPC+RMM-DIIS method.

For all algorithms tested in this section, we use the relative residual norm~\eqref{eq:rrn} as the stopping criteria and set the convergence tolerance to be $10^{-6}$.
To ensure a fair comparison with the SPPC, we use the eigenvectors of the zero-order part $H_0$ as the initial guesses for the algorithms.
All experiments were conducted using MATLAB.

%While the SPPC can make good eigenpair approximations in the early iterations, it can sometimes stagnate at later iterations.
%In such cases, we apply a hybrid approach, using the RMM-DIIS algorithm with the SPPC's approximations at the point of stagnation.
%We consider two types of numerical experiments: one for targeting the smallest eigenpair and the other for targeting the $k_{ev}$ smallest eigenpairs.

\subsection{Test Matrices}

We use the $A$-body Hamiltonian matrices corresponding to the nuclei ${}^6$Li, ${}^{7}$Li, ${}^{11}$B, and ${}^{12}$C in the following numerical experiments.
The superscripts indicate the number of nucleons in the nuclei; for example, ${}^7$Li indicates Lithium with 3 protons plus 4 neutrons.
The Hamiltonian matrices $H$ are constructed in a truncated CI space defined by a truncation parameter $\Nmax$, using the nucleon-nucleon interaction Daejeon16. 
For the same nucleus, a larger $\Nmax$ results in a larger matrix $H$, but the size of the matrix is independent of the interaction. 
Note that with three-nucleon interactions, the number of nonzero matrix elements is an order of magnitude larger than the one for the nucleon-nucleon interactions, and the number of iterations and the actual eigenvalues of any eigensolver will be different.
As we indicated earlier, the construction of $H$ can be done in a hierarchical fashion
so that a leading submatrix $\hat{H}_0$ of $H$ corresponds to the same nuclear $A$-body Hamiltonian represented in a lower dimensional configuration space associated with a smaller $\Nmax$.
In this section, if $H$ is the matrix representation of a nuclear $A$-body Hamiltonian represented in a configuration space associated with $\Nmax = n_c$, the submatrix $\hat{H}_0$ corresponds to the representation of the same Hamiltonian in a configuration space associated with $\Nmax = n_c - 2$.
We list the dimension of $H$ (denoted by dim($H$)), the dimension of $\hat{H}_0$ (denoted by dim($\hat{H}_0$)), the number of non-zero elements in $H$ (denoted by nnz($H$)), $\hat{H}_0$ (denoted by nnz($\hat{H}_0$)), and the $\Nmax$ value associated with $H$ in Table~\ref{tab:H}.
%and of $\hat{H}_0$ in Table~\ref{tab:H0}.
\begin{table}[b]
\caption{\label{tab:H}Properties of the test matrices $H$.}
\begin{ruledtabular}
\begin{tabular}{lcrrrr}
Nucleus    & $\Nmax$ & dim($H$) & dim($\hat{H}_0$) & nnz($H$)  & nnz($\hat{H}_0)$\\ \hline
${}^{6}$Li & 6  & 197,822  & 17,040 & 106,738,802 & 4,122,448 \\
${}^{7}$Li & 6  & 663,527  & 48,917 & 421,938,629 & 14,664,723 \\
${}^{11}$B & 4  & 814,092  & 16,097 & 389,033,682 & 2,977,735 \\
${}^{12}$C & 4  & 1,118,926 & 17,725 & 555,151,572 & 3,365,009  \end{tabular}
\end{ruledtabular}
\end{table}

%\begin{table}[b]
%\caption{\label{tab:H0}The leading submatrix $\hat{H}_0$ of the test matrix $H$.}
%\begin{ruledtabular}
%\begin{tabular}{llll}
%Nucleus    & $N_{\mbox{max}}$ & Matrix size $n_0$ & \# Non-zeros \\ \hline
%${}^{6}$Li & 4                & 17,040          & 4,122,448    \\
%${}^{7}$Li & 4                & 48,917          & 14,664,723   \\
%${}^{11}$B & 2                & 16,097          & 2,977,735    \\
%${}^{12}$C & 2                & 17,725          & 3,365,009   
%\end{tabular}
%\end{ruledtabular}
%\end{table}

\subsection{Targeting the Lowest Eigenpair}

We report the performance of the SPPC, the Lanczos algorithm, the LOBPCG algorithm, and the SPPC+RMM-DIIS for targeting the lowest eigenpair of the matrix $H$.
As mentioned earlier, we use a preconditioner for the LOBPCG and the RMM-DIIS algorithms.
The primary cost of all these algorithms is the number of SpMVs they perform before reaching convergence.
Because each algorithm performs one SpMV per iteration, we can directly compare them by the number of iterations required to reach convergence.

\begin{figure}
\includegraphics[scale=0.11]{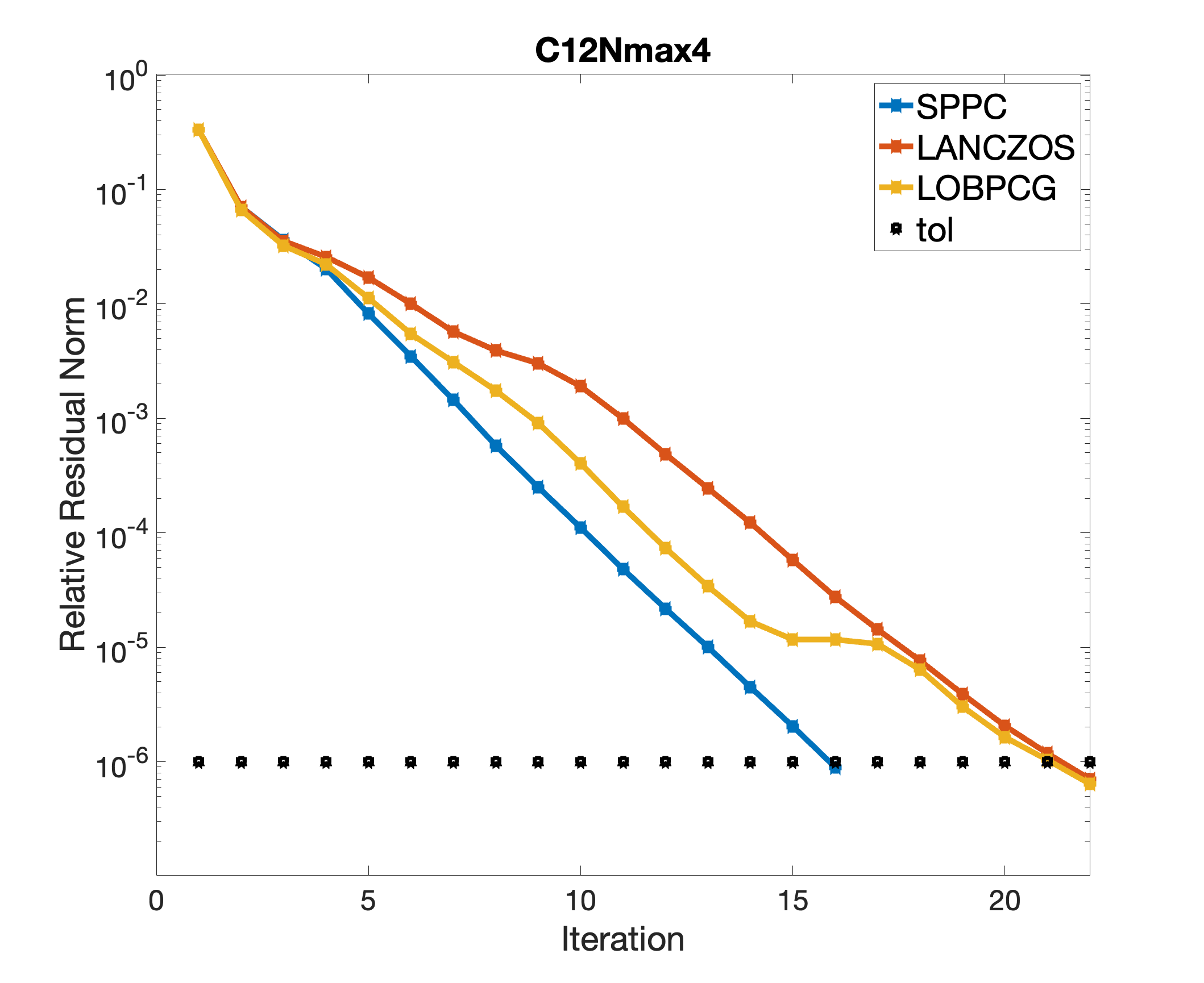}
\quad
\includegraphics[scale=0.11]{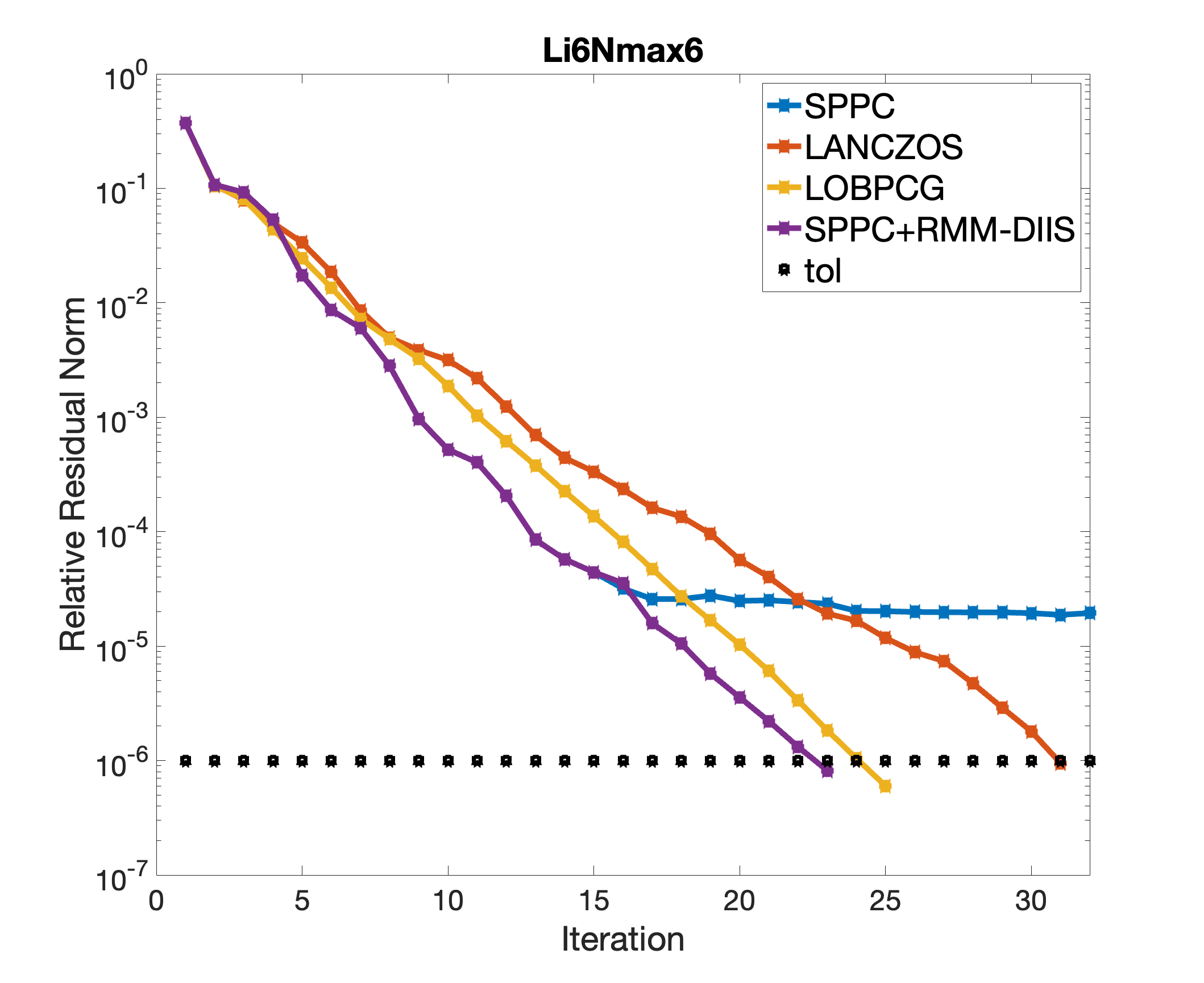}
\caption{\label{fig:ground}The convergence of the algorithms for computing the lowest eigenpair of the Hamiltonians matrices (${}^{12}$C on the left and ${}^{6}$Li on the right). One iteration equals one SpMV for all algorithms.}
\end{figure}

The left plot of Figure~\ref{fig:ground} shows the convergence history of the algorithms chosen for comparison for ${}^{12}$C with respect to the iteration number.
We observe that the SPPC algorithm converges in  $16$ iterations, which is the least among all methods, while the Lanczos algorithm and the LOBPCG algorithm converge in $22$ iterations.

\begin{figure}
\includegraphics[scale=0.11]{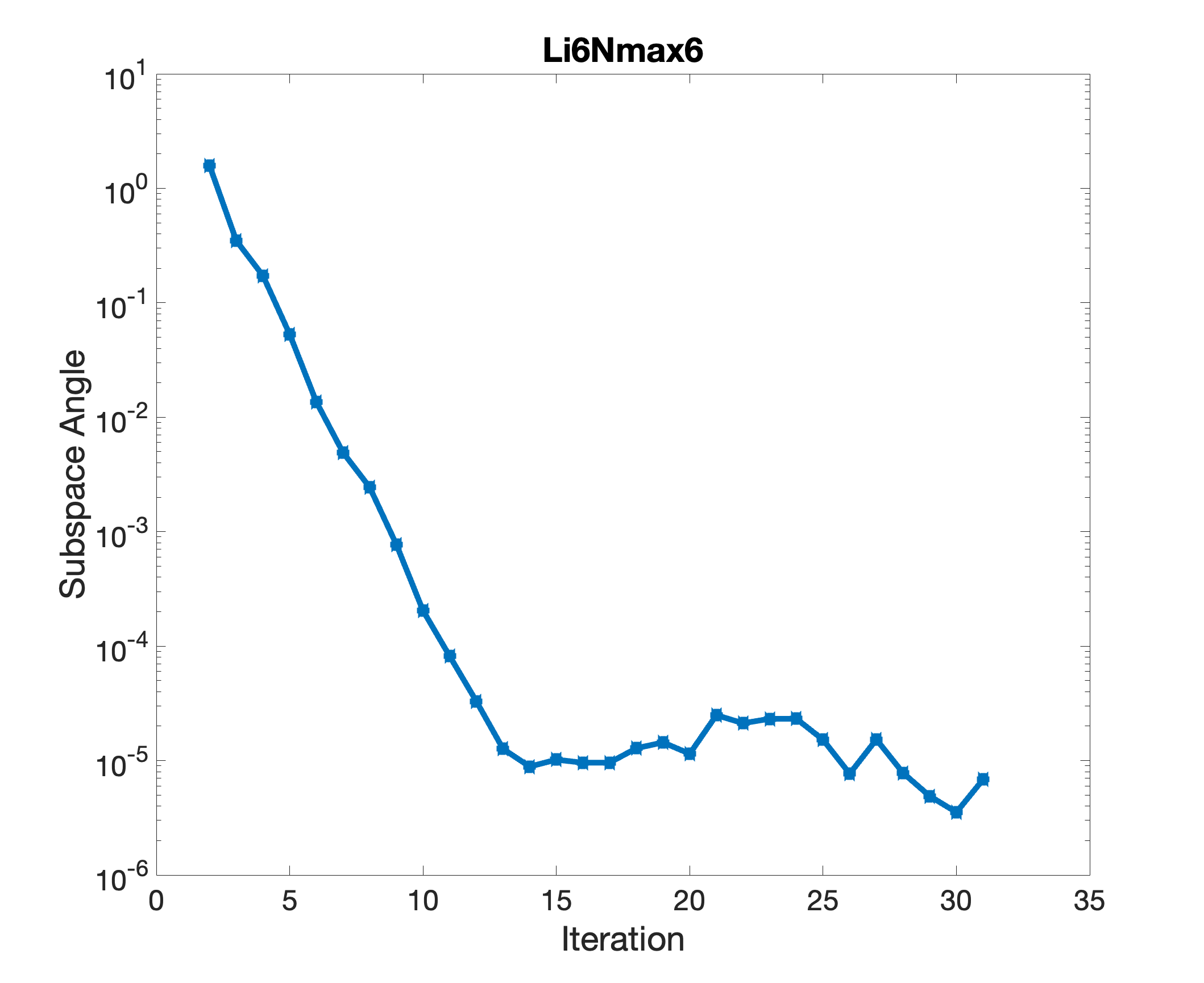}
\caption{\label{fig:subspace_angle}The subspace angle between the correction vector at iteration $p$ and the subspace spanned by the previous correction vectors from iteration $1$ to $p-1$.}
\end{figure}

The result shown in the right plot of Figure~\ref{fig:ground} is for the Hamiltonian associated with ${}^{6}$Li. Several features of the SPPC algorithm are observed.
The first observation is that the SPPC method converges more rapidly in the early iterations (up to the 15th iteration), and can be up to at most two orders of magnitude more accurate than other algorithms in these early iterations.
However, the SPPC approximation appears to stagnate in subsequent iterations.
This suggests that higher order correction vectors produced in later iterations (after iteration 15) do not contribute to improving the subspace constructed by the correction vectors produced in the early iterations.
To verify this conjecture, we plot the angle between the current correction vector and the subspace spanned by the previous correction vectors, denoted by $\angle(\Psi^{(p)},\mathcal{M}^{(p-1)})$, with respect to the iteration number $p$  in Figure~\ref{fig:subspace_angle}. We observe that $\angle(\Psi^{(p)},\mathcal{M}^{(p-1)})$ is relatively large in the first few SPPC iterations, and gradually decreases to the level of $10^{-5}$.  This is the point at which the new correction vector contributes minimally to the expansion of the subspace.

To overcome this stagnation, we consider a hybrid approach, the SPPC+RMM-DIIS, where we use the SPPC until the point of stagnation, and then switch to the RMM-DIIS.
This hybrid approach takes advantage of the fast convergence of the SPPC for the first few iterations and the fast convergence of the RMM-DIIS when initialized with a good initial guess to the desired eigenvector.
Specifically, we choose the initial guess in the RMM-DIIS as the Ritz vector produced from the SPPC method at the point of the switch.
For ${}^{6}$Li, we use the RMM-DIIS after the $15$th iteration of the SPPC.
We observe that the SPPC+RMM-DIIS breaks the stagnation of the SPPC and converges in $23$ iterations, while the Lanczos algorithm and the LOBPCG algorithm converge in $31$ and $25$ iterations, respectively.

Table~\ref{tab:SpMV_ground} gives a comparison of the SpMV counts used by several algorithms tested in this section for all four Hamiltonian matrices.
With the exception of the Hamiltonian of the nucleus ${}^{12}$C, the SPPC stagnates around iteration number 15. 
For these cases, we also consider the SPPC+RMM-DIIS.
We observe that this hybrid approach converges the fastest with the fewest SpMVs performed.
\begin{table}[b]
\caption{\label{tab:SpMV_ground}SpMV count of the algorithms for computing the lowest eigenpair. The convergence of the SPPC stagnates around 15 iterations for the Hamiltonians ${}^{6}$Li, ${}^{7}$Li, and ${}^{11}$B. For these three Hamiltonians, the hybrid method, the SPPC+RMM-DIIS, is also considered where the RMM-DIIS switches with the SPPC after iteration 15.}
\begin{ruledtabular}
\begin{tabular}{lllll}
Nucleus    & SPPC              & Lanczos & LOBPCG & SPPC+RMM-DIIS \\ \hline
${}^{6}$Li & \textgreater{}30 & 31      & 25     & 23           \\
${}^{7}$Li & \textgreater{}30 & 31      & 25     & 24           \\
${}^{11}$B & \textgreater{}30 & 30      & 27     & 18           \\
${}^{12}$C & 16               & 22      & 22     &             
\end{tabular}
\end{ruledtabular}
\end{table}

\subsection{Targeting the Five Lowest Eigenpairs}

\begin{figure*}
\includegraphics[scale=0.11]{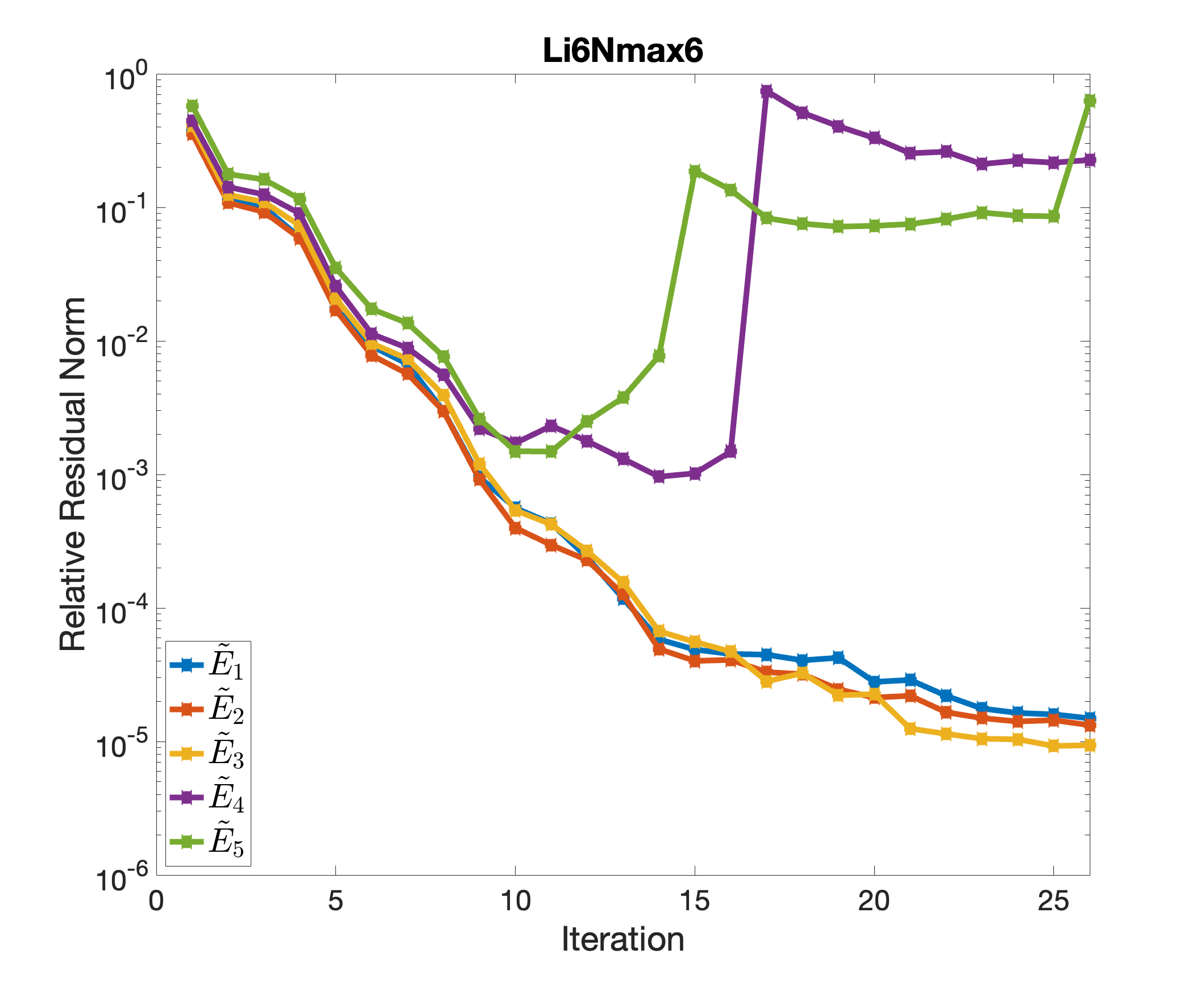}
\quad
\includegraphics[scale=0.11]{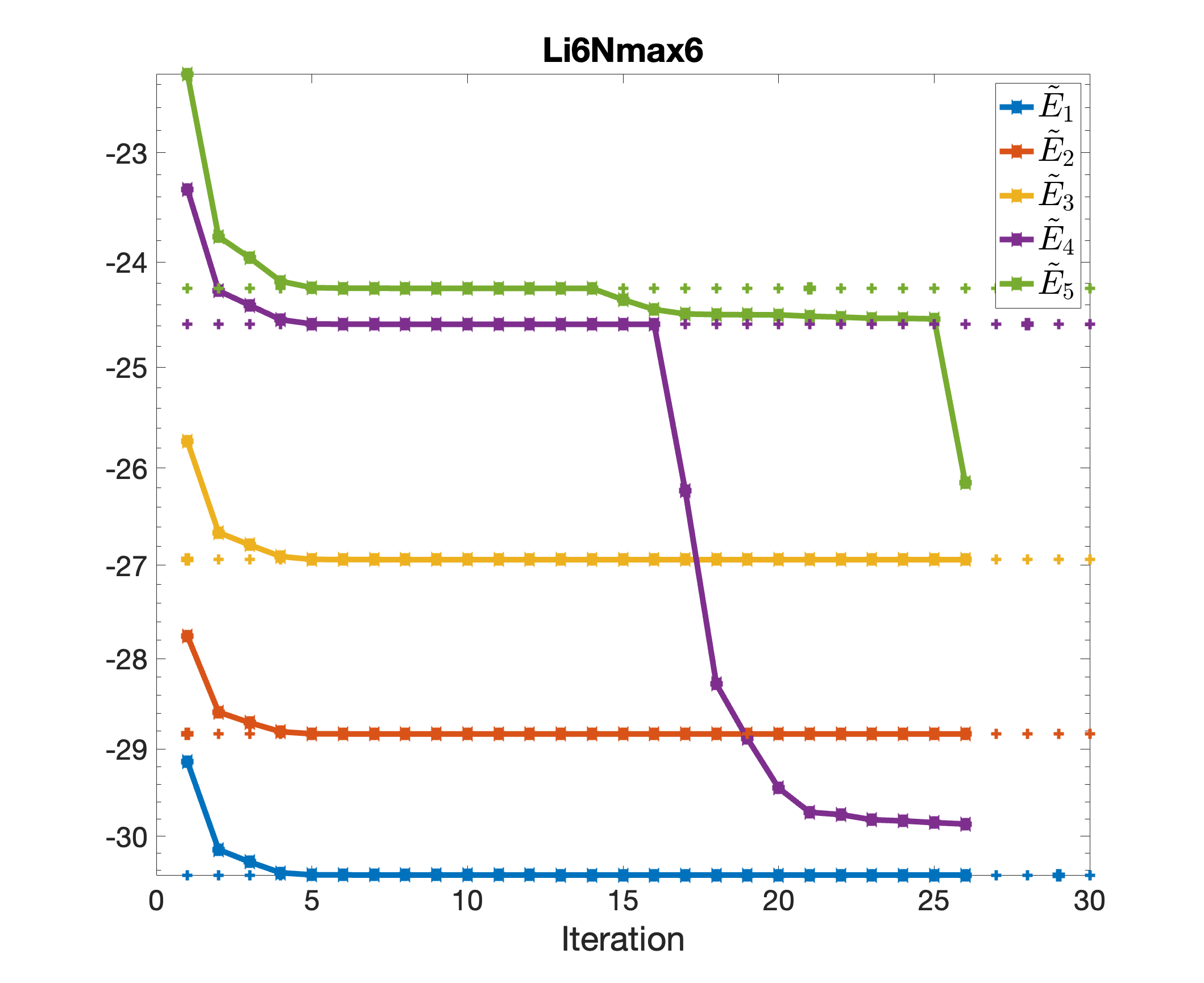}
\caption{\label{fig:SCCP_Li6Nmax6_Ind}Algorithm~\ref{alg:SPPC_k} to compute the first $5$ eigenpairs of the ${}^{6}$Li Hamiltonian, one by one. The left plot shows the relative residual norm, and the right plot shows the approximated eigenvalues in comparison with the true eigenvalues.}
\end{figure*}

\begin{figure*}
\includegraphics[scale=0.11]{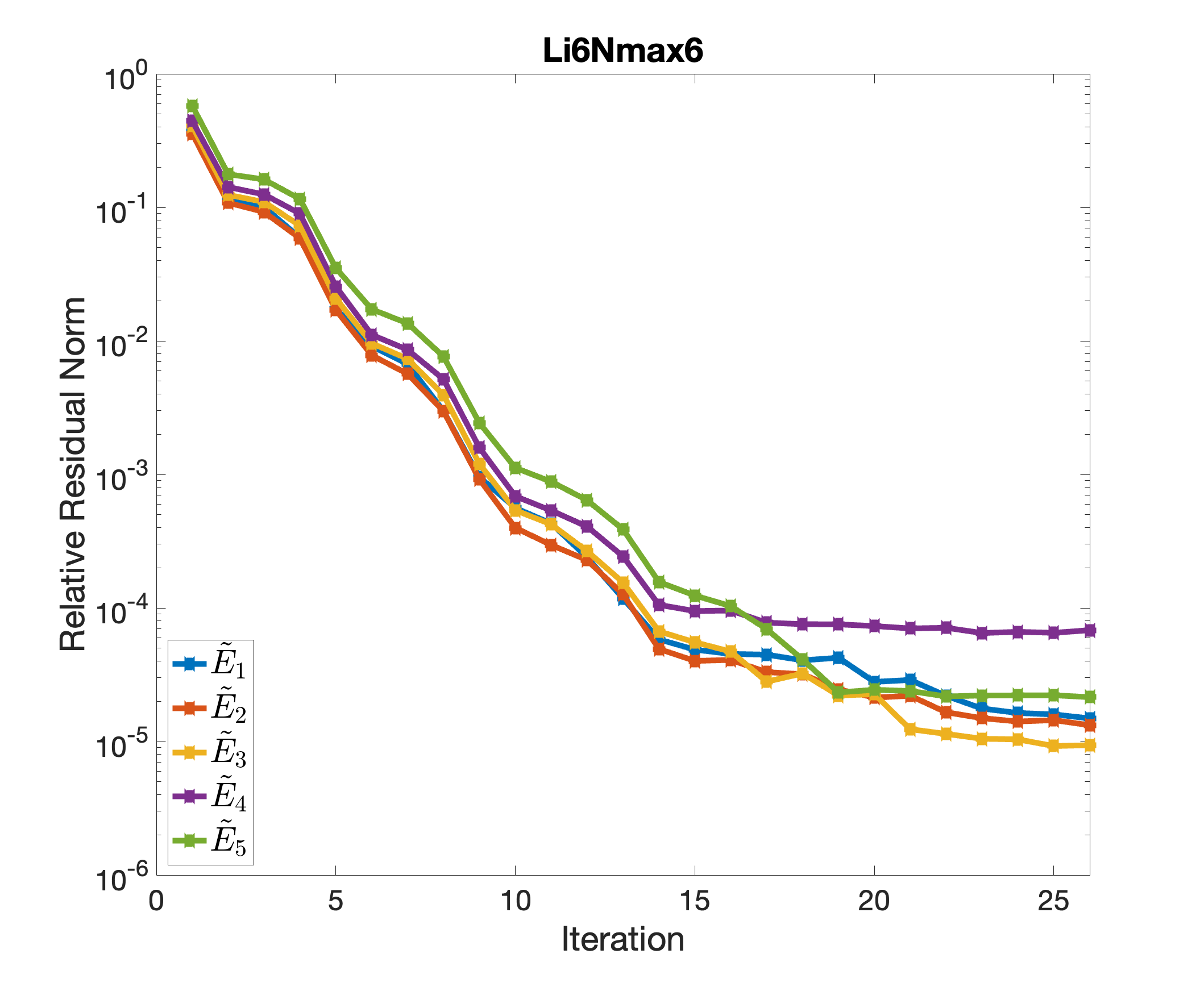}
\quad
\includegraphics[scale=0.11]{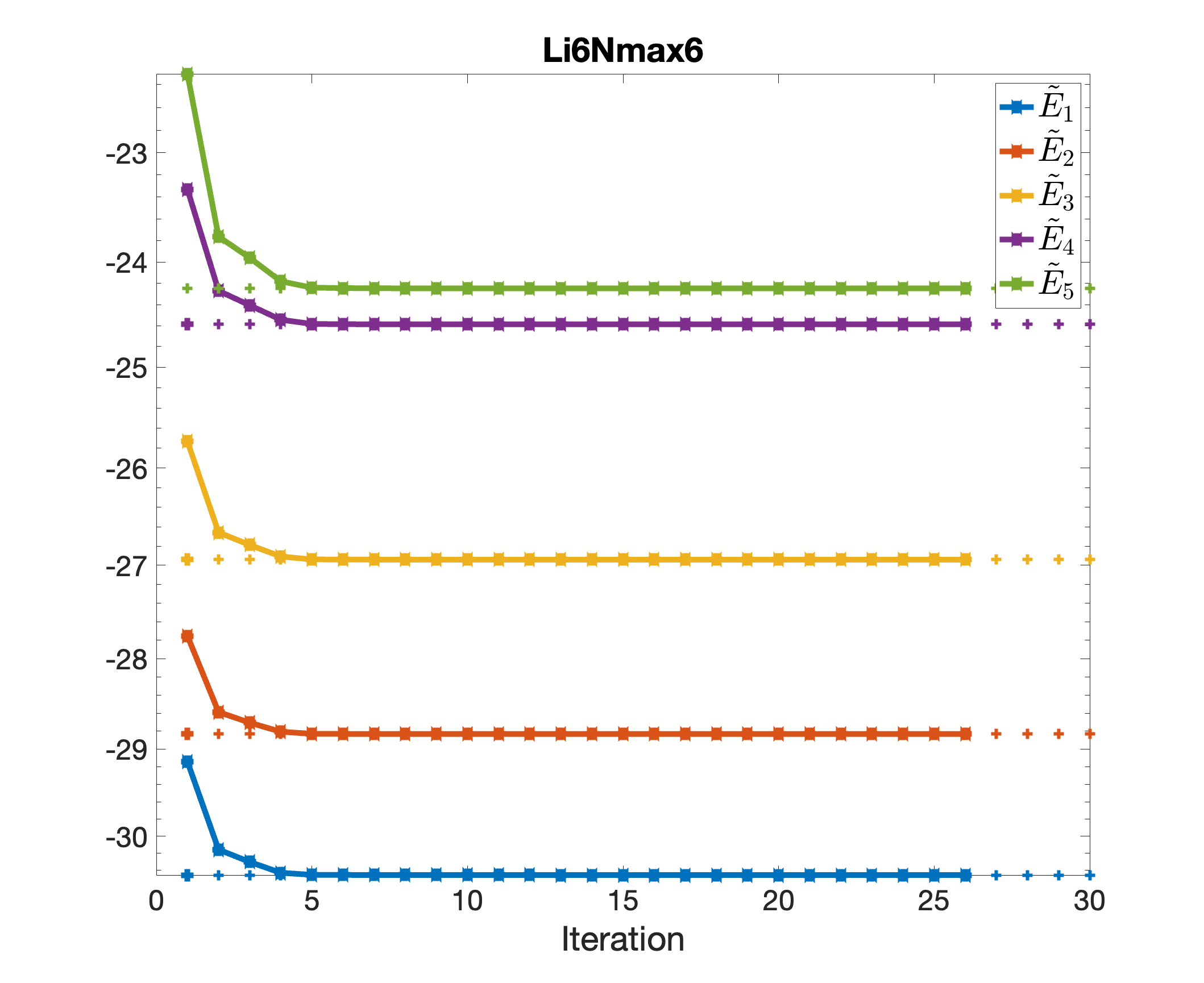}
\caption{\label{fig:SCCP_Li6Nmax6_Comb}Algorithm~\ref{alg:SPPC} to compute the first $5$ eigenpairs of the ${}^{6}$Li Hamiltonian. The left plot shows the relative residual norm, and the right plot shows the approximated eigenvalues in comparison with the true eigenvalues.}
\end{figure*}

We can use either Algorithm~\ref{alg:SPPC_k} or Algorithm~\ref{alg:SPPC} to compute the lowest $k_{ev}$ eigenvalues.
In the left plots of Figure~\ref{fig:SCCP_Li6Nmax6_Ind} and Figure~\ref{fig:SCCP_Li6Nmax6_Comb}, we show the relative residual norms of the approximation to the $5$ lowest eigenpairs of the ${}^{6}$Li Hamiltonian at each iteration of Algorithm~\ref{alg:SPPC_k} and Algorithm~\ref{alg:SPPC}, respectively.
The relative residual norms associated with the first three eigenpairs drop below $10^{-4}$ by the $15$th iteration.
However, for the fourth and the fifth eigenpairs, the relative residuals obtained by Algorithm~\ref{alg:SPPC_k} jump at a certain point and never become small in subsequent iterations.
The convergence failure for these eigenpairs can also be seen from  the change of Ritz values at each iteration and how they compare with the true eigenvalues of the Hamiltonian as shown in the right plot of Figure~\ref{fig:SCCP_Li6Nmax6_Ind}. It appears that the 4th and 5th Ritz values move towards a different eigenvalue.
This is due to a significant round off error in the orthogonalization process where the new basis contains contribution from the other eigenvector.
In contrast, as shown in the right plot of Figure~\ref{fig:SCCP_Li6Nmax6_Comb}, by constructing a larger subspace consisting of perturbative corrections to several eigenvectors, Algorithm~\ref{alg:SPPC} computes approximate eigenvalues that do not deviate from the true eigenvalues, and as a result, all of its computed eigenpairs converge within a reasonable accuracy.
Similar behaviors are observed for the other three Hamiltonians: ${}^{7}$Li, ${}^{11}$B, and ${}^{12}$C.

% \begin{figure*}
% \includegraphics[scale=0.11]{ECP-Ind_C12Nmax4_First5.png}
% \quad
% \includegraphics[scale=0.11]{ECP-Comb_C12Nmax4_First5.png}
% \caption{\label{fig:ECP_comp}Comparison between two approaches of the SPPC for targeting the smallest five eigenpairs. The left plot is for approach (a) and the right plot is for approach (b). $\lambda_k$ denotes the $k$th smallest eigenpair.}
% \end{figure*}

We now show that the stagnation in the convergence of the first Ritz pairs in the SPPC method can be eliminated by using a hybrid SPPC+RMM-DIIS. We switch to the RMM-DIIS method from the SPPC method when the relative residual norm of any of the eigenpairs starts to stagnate.
In this hybrid approach, a separate RMM-DIIS run is used to refine each approximate eigenpair. It is initialized with the corresponding Ritz vector returned from the SPPC method at the point of the switch.

\begin{figure}[b]
\includegraphics[scale=0.11]{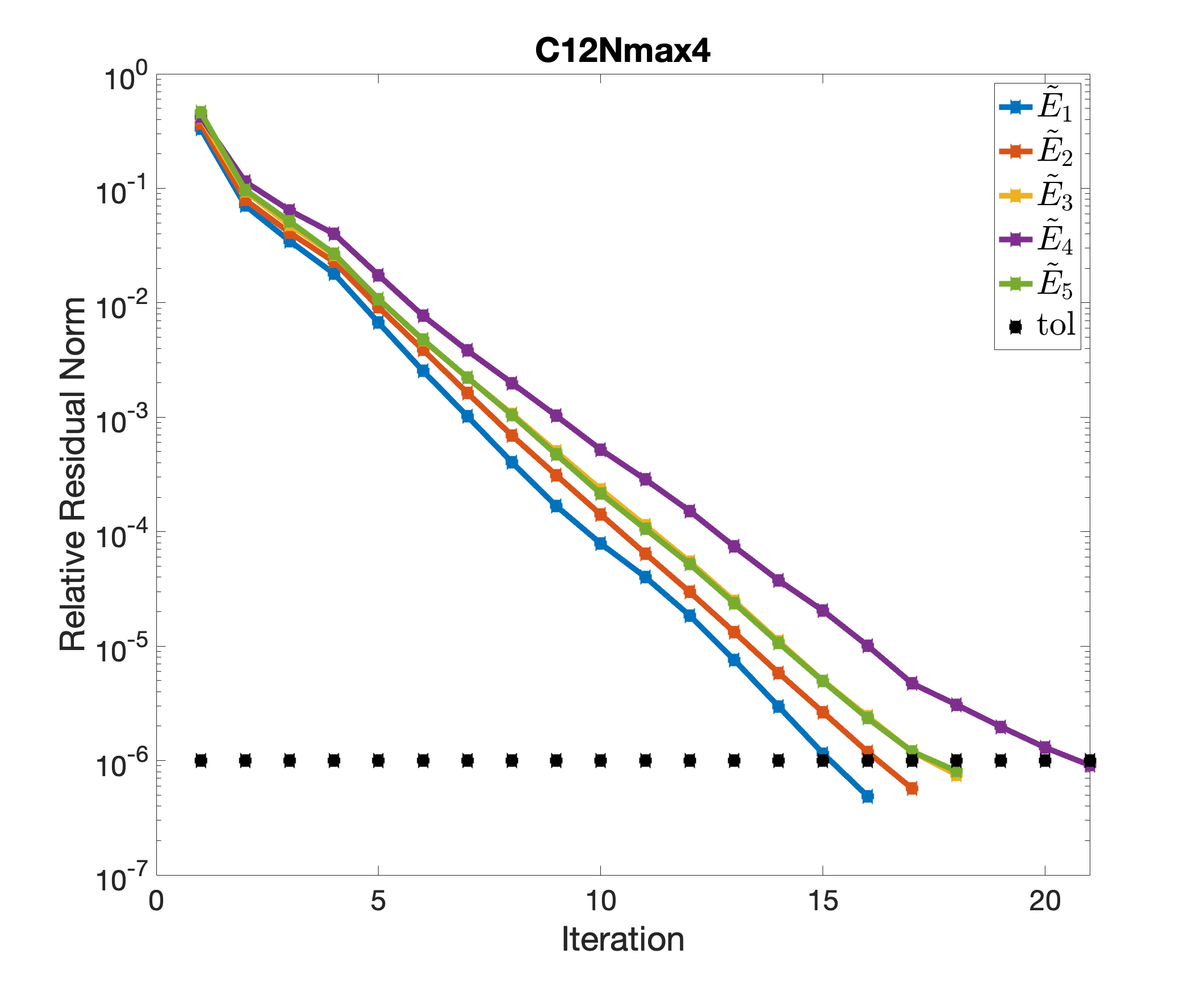}
\caption{\label{fig:hybrid_multi}The convergence of the SPPC+RMM-DIIS for computing the 5 lowest eigenapirs of the Hamiltonian ${}^{12}$C. The RMM-DIIS switches with the SPPC after iteration number 15. One iteration of an individual RMM-DIIS costs one SpMVs.}
\end{figure}

Figure~\ref{fig:hybrid_multi} illustrates the convergence of the SPPC+RMM-DIIS for the ${}^{12}$C Hamiltonian.
We choose to switch to the RMM-DIIS from the SPPC at iteration $15$, which is a point of stagnation for most eigenpairs.
We observe that the SPPC+RMM-DIIS breaks the stagnation of the SPPC and that all $5$ eigenpairs converge rapidly.

In Table~\ref{tab:multi}, we present the SpMV counts of the block Lanczos algorithm, the LOBPCG algorithm, and the SPPC+RMM-DIIS for computing the $5$ lowest eigenpairs of the four Hamiltonians.
The switch to the RMM-DIIS algorithm occurs for the SPPC+RMM-DIIS after the 15th iteration for all four Hamiltonian matrices.
The block Lanczos algorithm and the LOBPCG algorithm are block methods, thus requiring continued iterations until every eigenpair converges and consequently a full $k_{ev}$ SpMVs at each iteration.
In contrast, the SPPC+RMM-DIIS targets each eigenpair individually after it switches to RMM-DIIS, so it does not incur more SpMVs for the converged eigenpair as it targets the non-converged eigenpairs.
Due to this advantage, we observe that the SPPC+RMM-DIIS converges the fastest.
\begin{table}[b]
\caption{\label{tab:multi}SpMV count of the block Lanczos algorithm, the LOBPCG algorithm, and the SPPC+RMM-DIIS for computing the 5 lowest eigenpairs. The convergence of the SPPC tends to stagnate for some eigenpairs so the hybrid approach, the SPPC+RMM-DIIS, is considered. The RMM-DIIS switches from the SPPC after the 15th iteration.}
\begin{ruledtabular}
\begin{tabular}{llll}
Nucleus    & Block Lanczos & LOBPCG & SPPC+RMM-DIIS \\ \hline
${}^{6}$Li & 155           & 140    & 127          \\
${}^{7}$Li & 175           & 150    & 129          \\
${}^{11}$B & 170           & 125    & 94          \\
${}^{12}$C & 155           & 165    & 90         
\end{tabular}
\end{ruledtabular}
\end{table}

%%%%%%%%%%%%%%%%%%%%%%%%%%%%%%%%%%%%%%%%%%%%%%%%%%
\section{\label{sec:conclusion}Conclusion}

In conclusion, the Subspace Projection with Perturbative Corrections (SPPC) method, combined with the Residual Minimization Method with Direct Inversion of Iterative Subspace (RMM-DIIS), presents a significant advancement in the efficient computation of eigenpairs for large Hamiltonian matrices in nuclear structure calculations. 
The SPPC method leverages perturbative correction vectors to enhance the accuracy of eigenpair approximations in the initial iterations, substantially reducing the number of sparse matrix-vector multiplications (SpMVs) required for convergence. 
Although the SPPC may experience stagnation in subsequent iterations, this challenge is effectively mitigated by integrating it with the RMM-DIIS algorithm, which provides robust refinement of eigenvector approximations. 
Our numerical experiments across several nuclear Hamiltonians demonstrate that the SPPC+RMM-DIIS hybrid approach outperforms traditional methods in terms of SpMVs. 
This hybrid method offers a promising solution for large-scale nuclear structure calculations, providing a reliable and efficient approach to solving the $A$-body Schrödinger equation. 

While the preliminary results of the SPPC are promising, we have not provided a theoretical background explaining why it works. 
We plan to address this in our future work, discussing the convergence behavior in detail.
From a practical standpoint, w e aim to develop a method to automatically detect stagnation, eliminating the need for manual decisions on when to switch to the RMM-DIIS. 
Additionally, we are interested in implementing the SPPC in a hybrid MPI/OpenMPI code, such as the software MFDn (Many-Fermion Dynamics for nuclear structure) \cite{sternberg2008accelerating,maris2010scaling,aktulga2014improving}, to be run at high-performance computing centers. 
We also want to explore further optimizations and applications of the SPPC to other large-scale eigenvalue problems in nuclear physics and beyond.

%%%%%%%%%%%%%%%%%%%%%%%%%%%%%%%%%%%%%%%%%%%%%%%%%%
\begin{acknowledgments}
This material is based upon work supported by the Scientific Discovery through Advanced Computing (SciDAC) Program at the U.S. Department of Energy (DOE), Office of Science under Grants DE-SC0023495 and DE-SC0023175 from the Office of Nuclear Physics, and also under funding for the FASTMath Institute from the Office of Advanced Scientific Computing Research through Contract No. DE-AC02-05CH11231. This work used resources at the National Energy Research Scientific Computing Center (NERSC), which is funded by DOE under Contract No. DE-AC02-05CH11231. In addition, D.L. acknowledges partial support from DOE grants DE-SC0024586, DE-SC0023658, DE-SC0013365, and NSF grant PHY-2310620. J.P.V. acknowledges partial support from DOE grants DE-SC0023692 and DE-0023707.

% This material is based upon work supported by the U.S. Department of Energy, Office of Science, Office of Nuclear Physics under Grants DE-SC0023495 and DE-SC0023175, Office of Advanced Scientific Computing Research, Scientific Discovery through Advanced Computing (SciDAC) program via the FASTMath Institute under Contract No. DE-AC02-05CH11231. This work used resources of the National Energy Research Scientific Computing Center (NERSC) using NERSC Award ASCR-ERCAP m1027 for 2024, which is supported by the Office of Science of the U.S. Department of Energy under Contract No. DE-AC02-05CH11231.  D.L. also acknowledges partial support from grants DE-SC0024586, DE-SC0023658, DE-SC0013365, and NSF grant PHY-2310620. J.P.V. also acknowledges partial support from U.S. Department of Energy, Office of Nuclear Physics, under Grants DE-SC0023692 and DE-0023707. 

\end{acknowledgments}

\bibliography{apssamp}% Produces the bibliography via BibTeX.

\end{document}